\documentclass[a4paper,showpacs,showkeys,preprintnumbers,amsmath,amssymb,pra,twocolumn]{revtex4-1}

\usepackage[utf8]{inputenc}
\usepackage{amsopn}
\usepackage{bm,bbm}
\usepackage{amssymb,amsmath}
\usepackage{multirow}
\usepackage{hyperref}
\usepackage{graphicx}
\usepackage{subfig}
\usepackage{dsfont}
\usepackage{color}

\newcommand{\ket}[1]{\ensuremath{|#1\rangle}}
\newcommand{\bra}[1]{\ensuremath{\langle#1|}}
\newcommand{\ketbra}[2]{\ensuremath{\ket{#1}\bra{#2}}}
\newcommand{\braket}[2]{\ensuremath{\langle{#1}|{#2}\rangle}}
\newcommand{\HH}{\mathcal{H}}

\newcommand{\ee}{\mathrm{e}}
\newcommand{\ii}{\mathrm{i}}

\newcommand{\1}{{\rm 1\hspace{-0.9mm}l}}
\newcommand{\Id}{\1}

\newcommand{\ie}{\emph{i.e.\/}}

\newcommand{\C}{\ensuremath{\mathds{C}}}
\newcommand{\Cplx}{\ensuremath{\C}}

\begin{document}

\title{General model for a entanglement-enhanced composed quantum game on a
two-dimensional lattice}

\author{Jaros{\l}aw Adam Miszczak}
\author{{\L}ukasz Pawela}
\email{lukasz.pawela@gmail.com}
\affiliation{Institute of Theoretical and Applied Informatics, Polish Academy
of Sciences, Ba{\l}tycka 5, 44-100 Gliwice, Poland}

\author{Jan S{\l}adkowski}
\email{jan.sladkowski@us.edu.pl}
\affiliation{Institute of Physics, University of Silesia,
Uniwersytecka 4, 40-007 Katowice, Poland}

\begin{abstract}
We introduce a method of analyzing entanglement enhanced quantum games on
regular lattices of agents. Our method is valid for setups with periodic and
non-periodic boundary conditions. To demonstrate our approach we study two
different types games, namely the prisoner's dilemma game and a cooperative
Parrondo's game. In both cases we obtain results showing, that entanglement is
a crucial resource necessary for the agents to achieve positive capital gain.
\end{abstract}

\date{19/02/2014 (v. 0.20)}

\maketitle

\section{Introduction}
Systems involving a large number of simple variables with mutual interactions
appear frequently in various fields of research. Often the interactions are of
\emph{local character}. Network models have proven to be successful in analysis
of such phenomena \cite{Newman}. Axelrod \cite{axelrod81evolution,
axelrod06evolution} and Nowak and May \cite{nowak92evolutionary} prompted
scientist to investigate networks with local interactions that model various
social end economic structures. Cooperation and coordination are among the key
issues in economics and social sciences and can be analysed from this point of
view \cite{tomassini2007}. Game theoretical models often provide a qualitative
way of understanding various aspects of human decisions and behaviour of
complex systems. The combined network and game theoretical methods resulted in
a description of population structures with local interactions with sometimes
astonishing accuracy. The fields of econophysics and sociophysics were born
during the process \cite{RMS, Mirowski, Galam}. Bearing in mind interesting
analyses of a wide spectrum of problems performed from the quantum game
theoretical point of view \cite{Khrennikov, Busemeyer_2006, abbot,
pawela2013quantum}, one should not be astonished that quantum games have
invaded the network territory.

Quantum game theory approach extends such analyses in an interesting way
\cite{Busemeyer_2006, abbot, pawela2013quantum, giffen}. Cooperation is usually
modeled in the context of Prisoner's dilemma \cite{busemeyer1}. Another
interesting phenomenon is known as Parrondo paradox \cite{harmerabbot}, the
counterintuitive fact that in some cases combination of apparently loosing games
can result in success. Such games have their quantum counterparts and this
spurred us on to the analysis described in the present work.

The main contribution of this work is to provide a consistent model allowing to
study quantum games on two-dimensional lattice. We provide a general
ingredients needed in order to implement any quantum game in this general
scheme. In particular, we address the problem of using multipartite entangled
states shared by players. As a particular case, we study the family of games on
two-dimensional lattice constructed using the Parrondo scheme.

This paper is organized as follows. In Section \ref{sec:preliminaries} we
provide necessary background and notation. In Section~\ref{sec:model} we
develop a general model allowing the incorporation of a strategic quantum games
into the evolutionary scheme on two-dimensional graphs. In
Section~\ref{sec:examples} we use the introduced model to provide an uniform
analysis of a family of quantum games, focusing on the games based on the
Parrondo scheme. Finally, in Section~\ref{sec:conc} we draw the final
conclusions.

\section{Preliminaries}\label{sec:preliminaries}
Strategic game theory studies mathematical models of conflict and cooperation
between rational decision-makers \cite{osborne1994course} and is widely applied
in a great number of fields, ranging from biology to social sciences and
economics. Recently, the scientific community realized that quantum phenomena
might be important in this context. Therefore, a lot of attention has been given
to transferring concepts of game theory to the quantum realm hoping that this
work would contribute to our understanding of this difficult field of research. 

Quantum games are games in the standard sense but the approach allows for
harnessing quantum phenomena during the course of the game
\cite{piotrowski_invitation_2002, multiqubit_entangling}. The developed
formalism can be also used in a more abstract sense regardless of its the
quantum theoretical roots. Some classical game theoretical issues can be
extended to allow for quantum strategies including cooperation and coordination
problems. The set of quantum strategies is much larger than the set of
\emph{classical} ones and the presence of entanglement in the extended in such
way games ("quantized") implies more complex behavior of agents than the one
implied by "classical mixing" of strategies \cite{osborne1994course}.

For the purpose of this paper, $N$-player quantum game can be defined as a
quadruple
\begin{equation}
	\Gamma = (\HH,\rho,\mathcal{S},\mathcal{P}).
\end{equation}
In this ordered list $\HH$ is a Hilbert space, $\rho$ is a quantum state (i.e. a
density matrix), $\mathcal{S} = \{S_i\}_{i=1}^N$ is the set of possible player's
strategies and $\mathcal{P} = \{P_i\}_{i=1}^N$ is the set of payoff functions
for the players. Quantum strategies $s_i^\alpha \in S_i$ are completely positive
trace preserving (CPTP) maps. The payoff function of $i$-th player $P_i$ assigns
to a given strategy profile i. e. a set of player's strategies
$\{s_j^{\alpha_j}\}_{j=1}^N$ a real number -- the payoff.

Usually, the set of strategies is limited to unitary operators and the payoff is
determined via a measurement of the appropriate variables. A rich strategy sets
often allows for some spectacular results. For example, it has been shown that
if only one agent is aware of the quantum nature of the system, s/he will never
lose in some types of quantum games \cite{eisert1999quantum}. Moreover, it has
been demonstrated that a player can cheat by appending additional qubits to the
quantum system in question \cite{miszczak2011qubit}. One can also study the
impact of random strategies on the course of the game~\cite{kosik_quantum_2007}.

\section{Quantum games on 2D-lattice}\label{sec:model}
Let us consider a two-dimensional lattice with $M_1\times M_2$ nodes. Each node
of the lattice is occupied by one player.

Our scheme consists of three steps. In the first step, each agent in the network
is assigned an initial capital. In the second step, we create a quantum game
setup for a selected agent and all his/hers nearest neighbors. This referrers to
creation of an shared entangled state in standard Eisert quantization of games,
or a creation of a shared coin state in a quantum Parrondo game. After the
shared state is created, a quantum game is played by an agent with all of
his/her neighbors.

In the case of periodic boundary conditions, we need only to concern ourselves
with a five-player game, as every agent in a periodic lattice has four
neighbors. In the non-periodic case, we must also study the three- and
four-player game. Based on the results of the game and the game's payoff
function, a possible capital change vector is obtained.

In the third step, the capital values of each player is updated using the
capital change vectors from the previous step. This is done for every agent in
the lattice. This process is repeated $N$ times to enhance the differences in
capital behavior. This setup allows to study a number of different cases, as one
can assign different strategy setups for every number of players.

For a single player (node) on a lattice we introduce the following scheme for
playing entanglement assisted quantum game.
\begin{itemize}
\item The state $\ket{\psi_m}$ describing the game at node $m$ is used to
describe a subspace of coins and a subspace of position for each players,
\begin{equation}
\ket{\psi_m} \in \Cplx^{K_m^2} \otimes \Cplx^{K_m^4}
\end{equation}
where $K_m$ is the number of players playing with player at the node.

\item The initial state of the game is given as
\begin{equation}
\ket{\psi_m(0)}=\ket{\phi}\otimes \left( \bigotimes_{K_m} \ket{00} \right),
\end{equation}
where $\ket{\phi}$ is a multipartite state shared by $K_m$ players and the
position registers for all players are prepared in the base state.

\item Evolution operator is composed of the walker $W_m$ and the shift operator
$S_m$, where
\begin{equation}
\mathcal{W}=\bigotimes_{i=1}^{K_m} X(s_i) \otimes \Id^{K_m^4},
\end{equation}
with each $X(s_i) \in SU(2)$ and
\begin{equation}
\mathcal{S}=\sum_{i_1,\dots,i_{K_m}} \left(\bigotimes_{k=1}^{K_m} \ketbra{i_k}{i_k}
\right) \otimes \left( \bigotimes_{k=1}^{K_m} S^{-1^{k+1}}\right),
\end{equation}
where $S=\sum_{x}\ketbra{x+1}{x}$.
\end{itemize}

\section{Uniform analysis of entangled quantum games}\label{sec:examples}

\subsection{Prisoner's dilemma}
The quantum prisoner's dilemma game is defined as follows. Each player is sent 
a qubit and can locally operate
on it, using any unitary operator $U \in SU(2)$. The initial state of the 
system is entangled:
	\begin{equation}
		\ket{\psi} = J\ket{0}^{\otimes N},
	\end{equation}
where $N$ is the number of players, $J$ is the entangling 
operator~\cite{multiqubit_entangling}
	\begin{equation}
		J = \frac{1}{\sqrt{2}} \left(\Id^{\otimes N} + i\sigma_x^{\otimes 
		N}\right).
	\end{equation}
After the players have applied their respective strategies, the untangling 
gate, $J^\dagger$, is applied to the system,
hence the final state of the game is
	\begin{equation}
		\ket{\psi_f} = J^\dagger \left( \bigotimes_{i=1}^N U_i \right) 
		J\ket{0}^{\otimes N},
	\end{equation}
where $U_i$ is the strategy of the  $i$-th player.
The payoff of the first player amounts to:
	\begin{equation}
		\$_A = \sum_{i_1 \ldots i_N \in \{0,1\}^{\times N}} p_{i_1 \ldots i_N} 
		\braket{\psi_f}{i_1 \ldots i_N},
	\end{equation}
where $p_{i_1 \ldots i_N}$ are numbers corresponding to the possible classical 
payoffs of the first player. We assign the payoffs in the way 
described in~\cite{flitney2007nash}, Equation (23).

In the case of Prisoner's dilemma, the quantum version of the game utilizes an
extended set of strategies, which includes unitary operators. In the case of 
our scheme one can consider three scenarios by restricting the set of 
strategies available for the players, namely
\begin{enumerate}
\item only classical strategies, \ie\
\begin{equation}
\{C,D\},
\end{equation}
\item only quantum strategies from the set,
\begin{equation}
\{H,Q,\Sigma\},
\end{equation}
\item classical strategies with one quantum strategy,
\begin{equation}
\{ C,D, H \},\; \{C,D,Q\}, \; \{C,D,\Sigma\},
\end{equation}
\end{enumerate}
where the unitary strategies are
\begin{itemize}
    \item $C = \left(
        \begin{smallmatrix}
        1 & 0\\
        0 & 1
        \end{smallmatrix}
        \right)$,
    \item $D= \left(
        \begin{smallmatrix}
        0 & 1\\
        1 & 0
        \end{smallmatrix}
        \right)$,
    \item $H = 
        \frac{1}{\sqrt{2}}
        \left(
        \begin{smallmatrix}
        1 & 1\\
        1 &- 1
        \end{smallmatrix}
        \right)$,
    \item $Q =
        \left(
        \begin{smallmatrix}
        \ii & 0\\
        0 & -\ii
        \end{smallmatrix}
        \right)$,
    \item $\Sigma = 
        \left(
        \begin{smallmatrix}
        0 & 1\\
        -1 & 0
        \end{smallmatrix}
        \right).$
\end{itemize}
One should note that the strategy $H$ cannot be interpreted as a mixture of
classical strategies as it operates on the normalized vectors, whilst
the mixed strategy operates on the probabilities.

We perform 100 consecutive updates of the players' capital and we consider on
orthogonal neighbors. During each update, the players on a central node play
the game with five different players -- one in which they are at the center of
their four neighbors, and four where they are one of the neighbors for other
central players. Players on edge nodes play four games at each step -- one a
five-player game and the others four- or three-player games. A corner player
participates in two four-player games and one three-player game. In each of the
games a different set of entangled qubits is used.

\begin{figure}
	\includegraphics{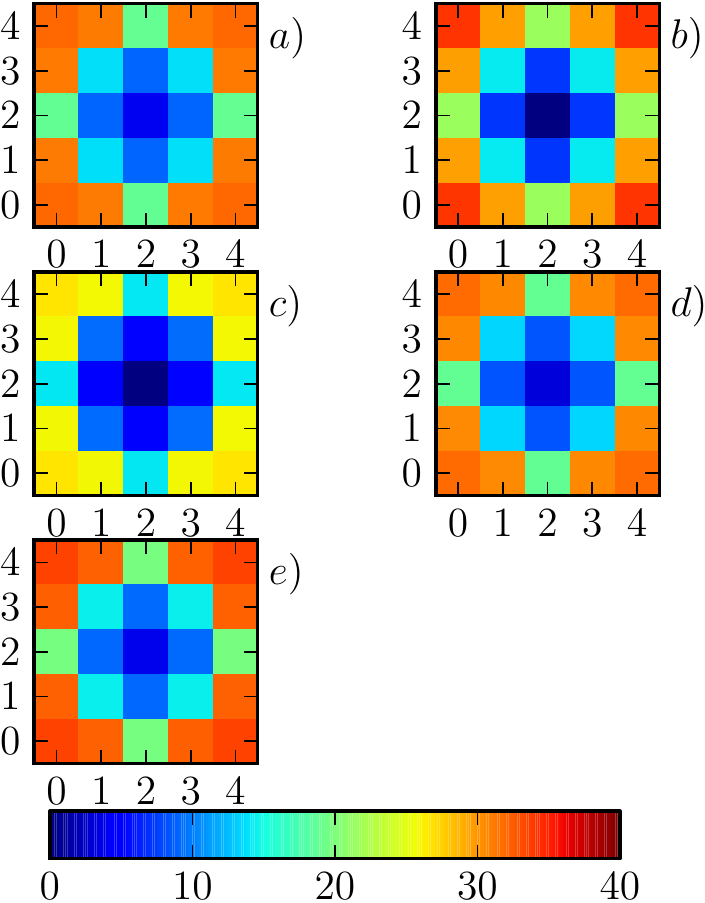}
	\caption{Average capital distribution for the prisoner's dilemma game in
	the case of classical strategies (a), quantum strategies (b), classical
	strategies with addition of the quantum strategies $H$, $Q$, $\Sigma$
	(c,d,e respectively).}\label{fig:pd_avg}
\end{figure}

\begin{figure}[!h]
\centering\includegraphics{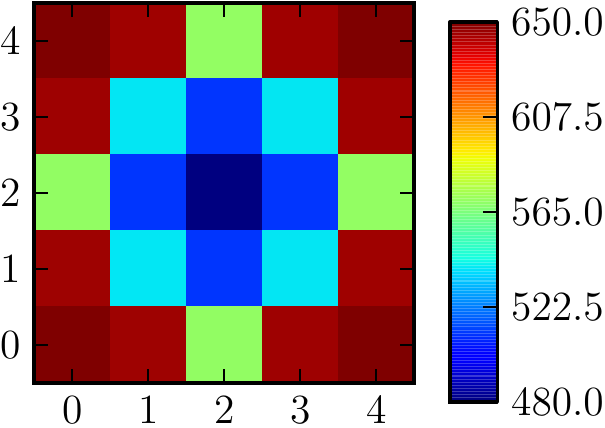}
\caption{Capital distribution for the prisoner's dilemma game in the case of
classical strategies. The strategy set is $[(C,C,C), (C,C,C,C), (C,C,C,C,C)]$
and gives the highest average capital of agents of all strategy
sets. This figure also illustrates the cases where we added a quantum strategy
to the classical ones.}\label{fig:pd_classical_best}
\end{figure}

The results obtained for the first case are shown in
Figure~\ref{fig:pd_avg}a. In it, we show the average capital of players
on a 5x5 network averaged over all possible strategy combinations. As can be
seen, the highest average capital gains are achieved by players closer to the
edge of the lattice. The average capital of an agent in the lattice is
equal to 19. The strategy set which yields the highest average
capital of a player is $[(C,C,C), (C,C,C,C), (C,C,C,C,C)]$ and gives an
average capital of 588.8. A strategy set $(A, B, C)$ means 
that the first player uses strategy $A$, the second strategy $B$ and the third 
strategy $C$. The capital distribution for this case is shown in
Figure~\ref{fig:pd_classical_best}

\begin{figure}[!h]
\centering\includegraphics{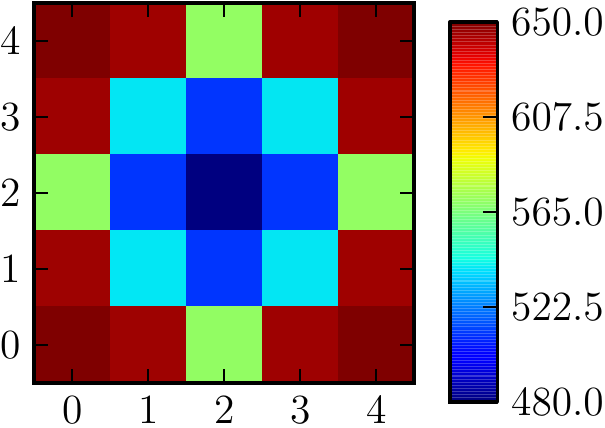}
\caption{Capital distribution for the prisoner's dilemma game in the case of
quantum strategies. The strategy set is $[(\Sigma,\Sigma,\Sigma), (Q,Q,Q,Q),
(\Sigma,\Sigma,\Sigma,\Sigma,\Sigma)]$ and gives the highest average capital of
agents of all strategy sets.}\label{fig:pd_quantum_best}
\end{figure}

The second case is illustrated in Figure~\ref{fig:pd_avg}b. In this case,
similar to the classical one, highest capital gains are found for the edge
players. Also, the overall capital distribution is the same as in the purely
classical case. The average capital of an agent in the lattice is equal to
$17.8$. The strategy set which yields the highest average capital of a player
is $[(\Sigma,\Sigma,\Sigma), (Q,Q,Q,Q), (\Sigma,\Sigma,\Sigma,\Sigma,\Sigma)]$
and gives an average capital of 588.8. The capital distribution for this case
is shown in Figure~\ref{fig:pd_quantum_best}

We present the results for the third case in
Figures~\ref{fig:pd_avg}c, \ref{fig:pd_avg}d and
\ref{fig:pd_avg}e for the possible strategies $\{C,D,H\}$,
$\{C,D,Q\}$ and $\{C,D,\Sigma\}$ respectively. The average capital gains are
-29.6, 15.4 and 28.2 respectively. As can be seen, the average capital gain
with the added strategy $\Sigma$ is higher than in the purely classical case.
The strategy set which gives the highest average capital gain of a player is
$[(C,C,C), (C,C,C,C), (C,C,C,C,C)]$, the same as in the purely classical case.
Hence, the capital distribution for this case is identical to that shown in
Figure~\ref{fig:pd_classical_best} and the average capital gain is equal to
588.8.

\subsection{Cooperative Parrondo paradox}
Cooperative Parrondo's games were introduced by Toral
\cite{toral_cooperative_2001}. The scheme is as follows. Consider an ensemble of
$N$ players, each with his/hers own capital $C_i(t)$, $i = 1, 2, \ldots, N$. As
in the original paradox, we consider two games, A and B. Player $i$ can play
either game A or B according to some rules. The main difference from the
original paradox is that probabilities of game B depend, in general, on the
state of all players $j \neq u$. For simplicity, we only consider the case when
the probabilities of winning at time $t$, depend only on the present state of
the players. The game, by definition, is a winning one, when the average value
of the capital
\begin{equation}
\langle C(t) \rangle = \frac1N \sum_{i=1}^N C_i(t),\label{eq:avpay}
\end{equation}
increases with time.

There are several known approaches to quantization of Parrondo's
games~\cite{flitney_quantum_2002,gawron_quantum_2005}. We model a cooperative
quantum Parrondo's game as a multidimensional quantum walk
(QW)~\cite{flitney_quantum_2004,chandrashekar2011parrondos}. The average
position of the walker along each axis determines each player's payoff. As in
the classical case, we consider two games, $A$ and $B$. Similar to the
classical case the probabilities of winning game $B$ depend on the state of the
other players. A detailed introduction of the model of the game is presented
in~\cite{pawela2013cooperative}.

The following two
possible schemes of alternating between games $A$ and $B$ are considered
\begin{enumerate}
	\item random alternation, denoted $A+B$
	\item games played in succession $AABBAABB\ldots$, denoted $[2,2]$.
\end{enumerate}

We focus our attention on the cooperative game in a regular lattice. We consider
two types of boundary conditions for the lattice: periodic, allowing to simulate
the behavior of an infinite lattice, and non-periodic to simulate the finite
case.

With each agent on the network we associate a Hilbert space that consists of two
components: the coin's Hilbert space and the position Hilbert space
\begin{equation}
	\HH_i = \HH_c \otimes \HH_{pos}.
\end{equation}

We introduce two base states in the single coin Hilbert space, the $\ket{L}$
and
$\ket{R}$ states. These states
represent the classical coin's heads and tails respectively.

When the agents play the game, we connect their respective Hilbert spaces using
the tensor product
\begin{equation}
	\HH_G = \bigotimes_{i=1}^N \HH_i,
\end{equation}
where $N$ denotes the total number of players participating in the game.

Game A is implemented using an operator performing a flip of a fair coin. The
operator is given by
\begin{equation}
U_A = \left(
		\begin{array}{cc}
			\frac{1}{\sqrt{2}} & \frac{\ii}{\sqrt{2}} \\
			\frac{\ii}{\sqrt{2}} & \frac{1}{\sqrt{2}}
		\end{array}
	\right).
\end{equation}

The game B is played as follows. An agent $i$ performs a flip of his coin using
one of the unitary operators from the family
\begin{equation}
	U_k = \left(
		\begin{array}{cc}
			\sqrt{\rho_k} & \sqrt{1 - \rho_k}\ee^{\ii\theta_k} \\
			\sqrt{1 - \rho_k}\ee^{\ii\phi_k} & -\sqrt{\rho_k}\ee^{\ii(\theta_k
			+
			\phi_k)}
		\end{array}
	\right).
\end{equation}
Each operator in the set $\{U_k\}$ depends on the parameters: $\rho_k$ which is
the classical probability that the coin does not change its state,
and $\phi_k$ and $\theta_k$ are phase angles, which we assume to be $\phi_k =
\theta_k = \pi/2$ for all $k$. The choice of the probabilities, and thus an
operator from the set $\{U_k\}$, depends on the number of neighbors and the
number of winners and losers amongst the neighbors of an agent $i$.

We set the probabilities in game B to $\rho=0.5$ except for 
the case when all the other players have lost. In this case, we set it to 
$\rho=0.9$.

After the coin flip, agent $i$ applies a move operator to his $\HH_i$ space.
The operator is given by
\begin{equation}
	U_{pos} = P_r \otimes S + P_l \otimes S^\dagger,
\end{equation}
where $S$ is a shift operator in the position space, defined as $S\ket{x} =
\ket{x+1}$, $\ket{x}$ denotes the current position of the walker. It follows
from the definition of $S$ that $S^\dagger\ket{x} = \ket{x-1}$. $P_r$ and $P_l$
denote projection operators on the $\ket{R}$ and $\ket{L}$ states of the coin
respectively.

We study the behavior of the lattice for different sets $\{U_k\}$ of possible
game B coin tosses. We chose those sets so that
\begin{enumerate}
	\item a three and four player game shows paradoxical behavior,
	\item game shows paradoxical behavior for three, four and five players
\end{enumerate}
Furthermore, we study different initial states of players' coins: the separable
state, the entangled GHZ state and the entangled W state.

The setup of the simulation is as follows. Each agent in the network plays a
game with all of his/hers neighbors. The network is updated sequentially. The
network is evolved for 1000 iterations. In the case of the A+B game scheme, we
average the results for 10 independent runs of the simulation.

\begin{figure}[h]
	\subfloat[{[2,2]} non-periodic lattice.]
	{\label{fig:separable_0}\includegraphics{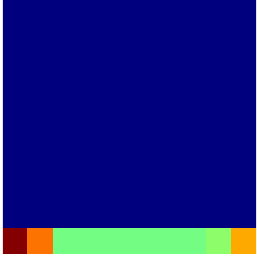}}
	\hspace{0.3cm}
	\subfloat[A+B non-periodic lattice.]
	{\label{fig:separable_1}\includegraphics{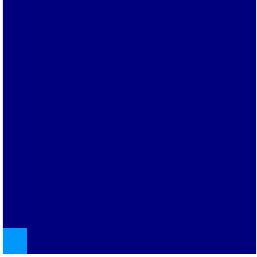}}\\
	\subfloat[{[2,2]} periodic lattice.]
	{\label{fig:separable_2}\includegraphics{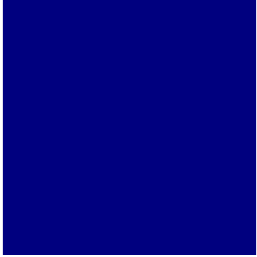}}
	\hspace{0.3cm}
	\subfloat[A+B periodic lattice.]
	{\label{fig:separable_3}\includegraphics{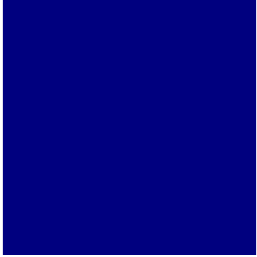}}\\
	\subfloat{\includegraphics{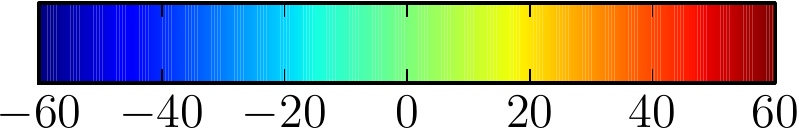}}
	\setcounter{subfigure}{4}
	\subfloat[Average payoff as a function of the number of players. The bars 
	show the payoff for game A, game B, game {[2,2]} and game A+B going from 
	left to right.]
	{\label{fig:separable_effect}\includegraphics[width=0.49\textwidth]{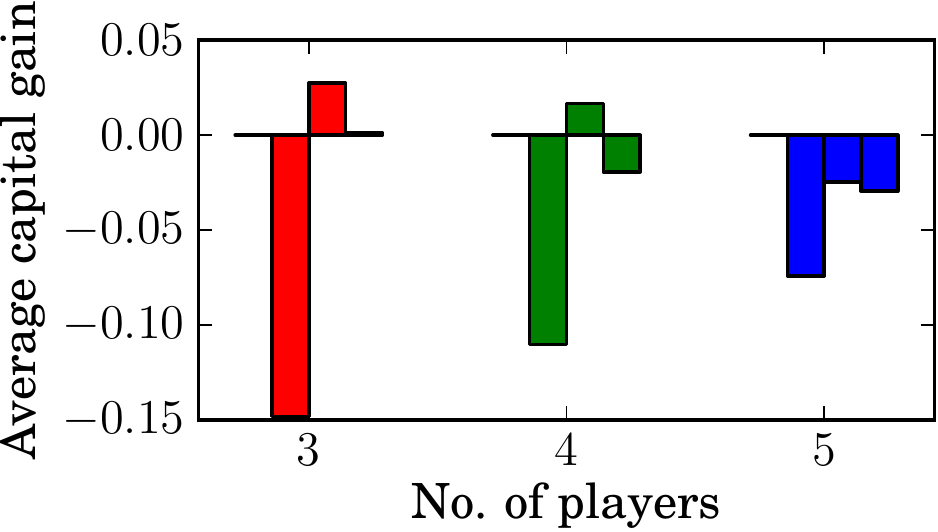}}\\
	\subfloat[The average capital of the networks shown in 
	Figures~\ref{fig:separable_0}, \ref{fig:separable_1}, 
	\ref{fig:separable_2}, \ref{fig:separable_3}]
	{\label{fig:separable_avg_cap}\includegraphics{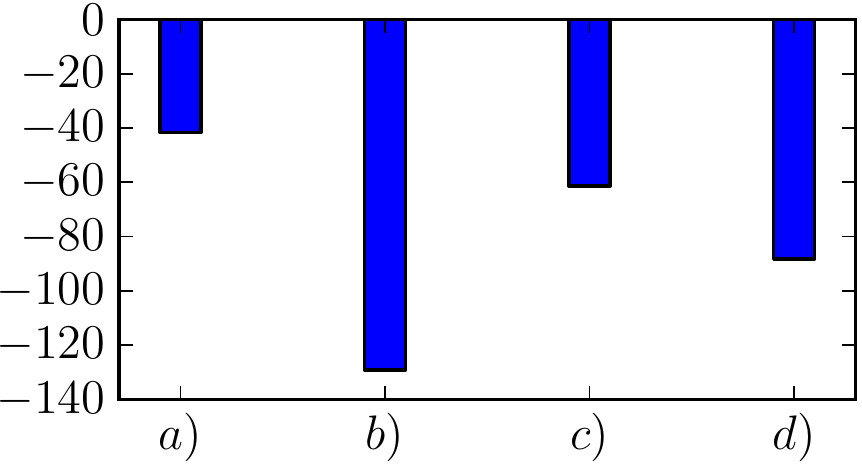}}
	\caption{The state of the network for the [2,2] game on non-periodic
	lattice (a)), A+B game on non-periodic lattice (b)), [2,2] game on
	periodic lattice (c)) and A+B game on periodic lattice (d)) after 1000
	iterations. The initial coin state is the separable state. Figure e) shows
	the payoff for games A, B, [2,2] and A+B as a function of the number
	of players. Figure f) shows the average capital gain of an agent in
	the network after 1000 iterations.}\label{fig:separable}
\end{figure}

Figure~\ref{fig:separable} shows the results of the simulation when the initial
state of the coin is separable. Figures~\ref{fig:separable_0} and
\ref{fig:separable_1} show the results of the simulation for the non-periodic
network and Figures~\ref{fig:separable_2} and \ref{fig:separable_3} show the 
results
for the periodic network. In the case of non-periodic network, the final
structure of the network appears similar for both games: the only
difference being the absolute value of capital of each agent. The same is true
for the periodic network. Bar plots presented in 
Figure~\ref{fig:separable_effect} shows
the average capital gains of the players in a three, four and five-player
game. The game A+B shows paradoxical behavior only in the three-player case,
whereas the [2,2] game shows this kind of behavior in three and four-player
games. As the five-player game is always a losing one and the
network contains mostly agents playing the five player game, the average
capital gain of an agent in the network is negative. In fact, almost every
agent experiences a capital decrease. Figure~\ref{fig:separable_avg_cap} shows 
the average capital of an agent in all studied cases.

\begin{figure}[h]
	\subfloat[{[2,2]} non-periodic lattice.]
	{\label{fig:GHZ_0}\includegraphics{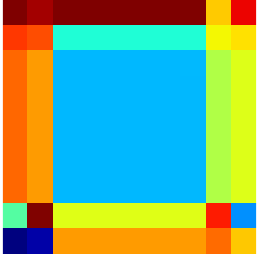}}
	\hspace{0.3cm}
	\subfloat[A+B non-periodic lattice.]
	{\label{fig:GHZ_1}\includegraphics{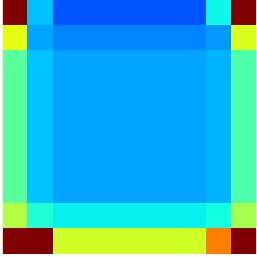}}\\
	\subfloat[{[2,2]} periodic lattice.]
	{\label{fig:GHZ_2}\includegraphics{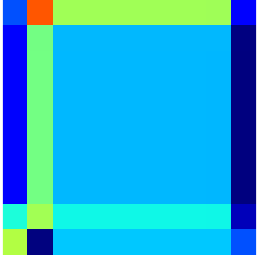}}
	\hspace{0.3cm}
	\subfloat[A+B periodic lattice.]
	{\label{fig:GHZ_3}\includegraphics{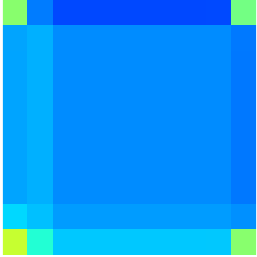}}\\
	\subfloat{\includegraphics{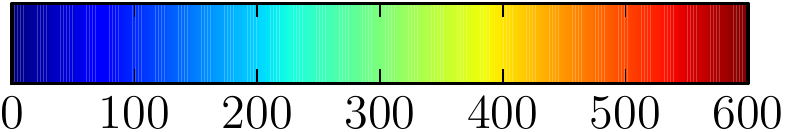}}
	\setcounter{subfigure}{4}
	\subfloat[Average payoff as a function of the number of players. The bars 
	show the payoff for game A, game B, game {[2,2]} and game A+B going from 
	left to right.]
	{\label{fig:GHZ_effect}\includegraphics[width=0.49\textwidth]{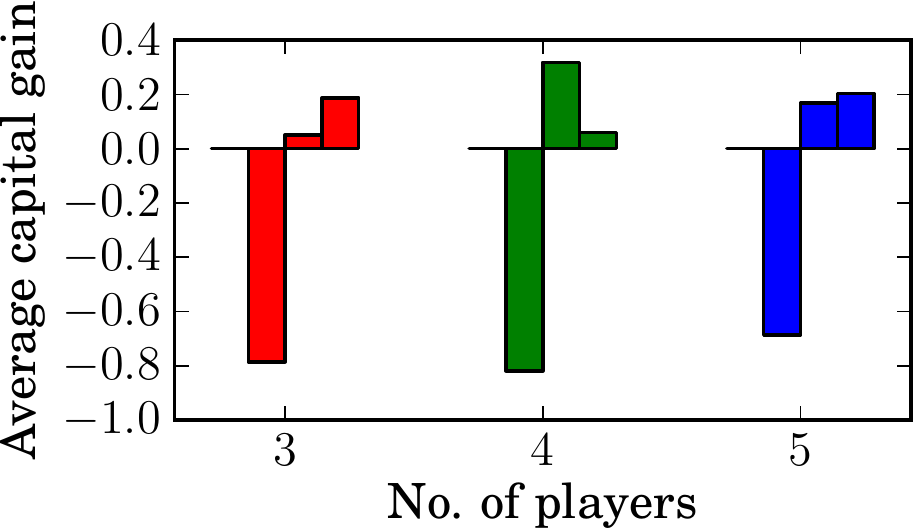}}\\
	\subfloat[The average capital of the networks shown in 
	Figures~\ref{fig:GHZ_0}, \ref{fig:GHZ_1}, \ref{fig:GHZ_2}, \ref{fig:GHZ_3}]
	{\label{fig:GHZ_avg_cap}\includegraphics{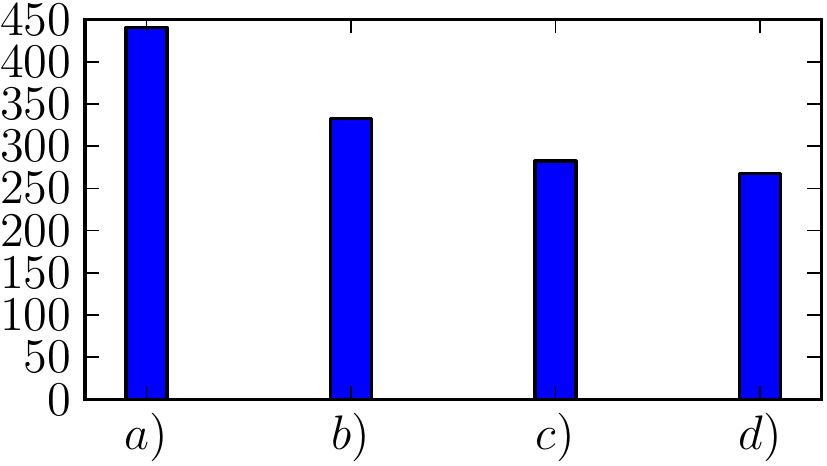}}
	\caption{The state of the network for the [2,2] game on non-periodic
	lattice (a)), A+B game on non-periodic lattice (b)), [2,2] game on
	periodic lattice (c)) and A+B game on periodic lattice (d)) after 1000
	iterations. The initial coin state is the GHZ state. Figure e) shows
	the payoff for games A, B, [2,2] and A+B as a function of the number
	of players. Figure f) shows the average capital gain of an agent in
	the network after 1000 iterations.}\label{fig:GHZ}
\end{figure}

In the case of the GHZ state, the results are summarized in
Figure~\ref{fig:GHZ}. Figures~\ref{fig:GHZ_0} and  \ref{fig:GHZ_1} show the
results of the simulation for the non-periodic  network and
Figures~\ref{fig:GHZ_2} and \ref{fig:GHZ_3} show the results  for the periodic
network. The capital distribution in the network differs significantly from the
one obtained for the separable coin state. In this case, games  [2,2]  and A+B
show paradoxical behavior in the case of three, four and five players as
depicted in Figure~\ref{fig:GHZ_effect}. Also, the value of average capital
gain in each of these games is much higher than  in  the case of the separable
coin state. This is reflected in the  average capital  gain of the entire
network after 1000 iterations. As shown in Figure~\ref{fig:GHZ_avg_cap} the
average capital gain of the network is much greater than in the separable case.
This is due to the fact, that all games show paradoxical behavior. Therefore,
the agents in the network always gain capital.

\begin{figure}[h]
	\subfloat[{[2,2]} non-periodic lattice.]
	{\label{fig:W_0}\includegraphics{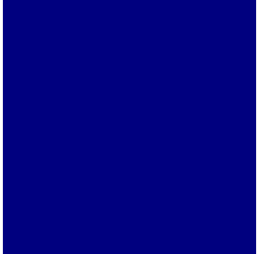}}
	\hspace{0.3cm}
	\subfloat[A+B non-periodic lattice.]
	{\label{fig:W_1}\includegraphics{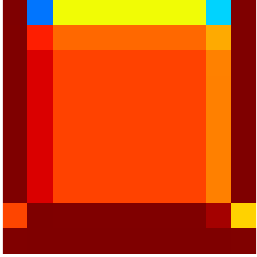}}\\
	\subfloat[{[2,2]} periodic lattice.]
	{\label{fig:W_2}\includegraphics{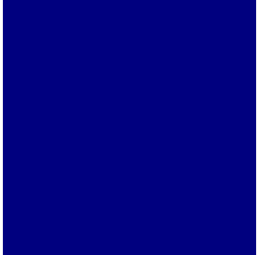}}
	\hspace{0.3cm}
	\subfloat[A+B periodic lattice.]
	{\label{fig:W_3}\includegraphics{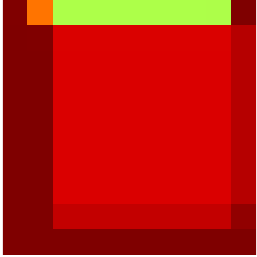}}\\
	\subfloat{\includegraphics{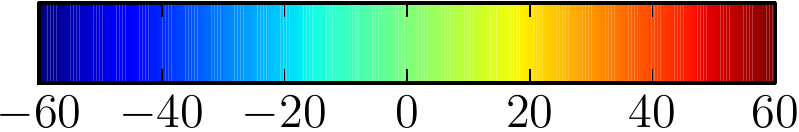}}
	\setcounter{subfigure}{4}
	\subfloat[Average payoff as a function of the number of players. The bars 
	show the payoff for game A, game B, game {[2,2]} and game A+B going from 
	left to right.]
	{\label{fig:W_effect}\includegraphics[width=0.49\textwidth]{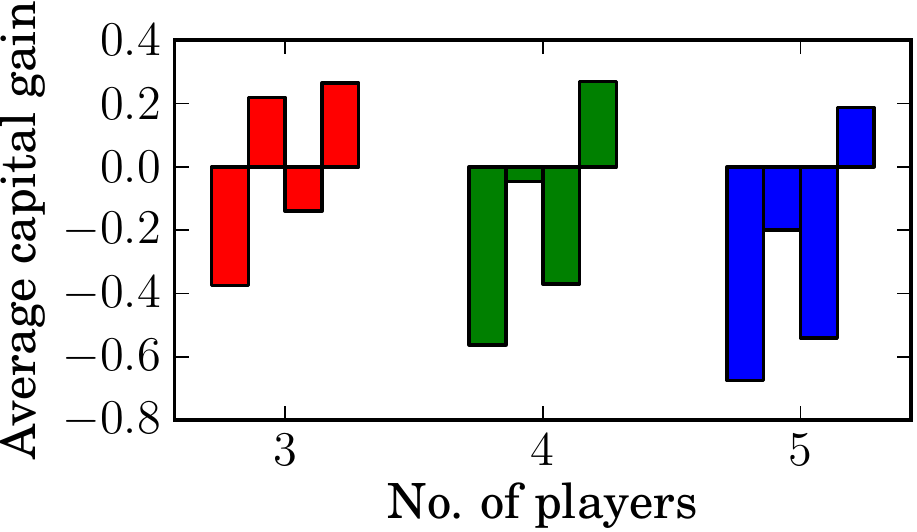}}\\
	\subfloat[The average capital of the networks shown in 
	Figures~\ref{fig:W_0}, \ref{fig:W_1}, \ref{fig:W_2}, \ref{fig:W_3}]
	{\label{fig:W_avg_cap}\includegraphics{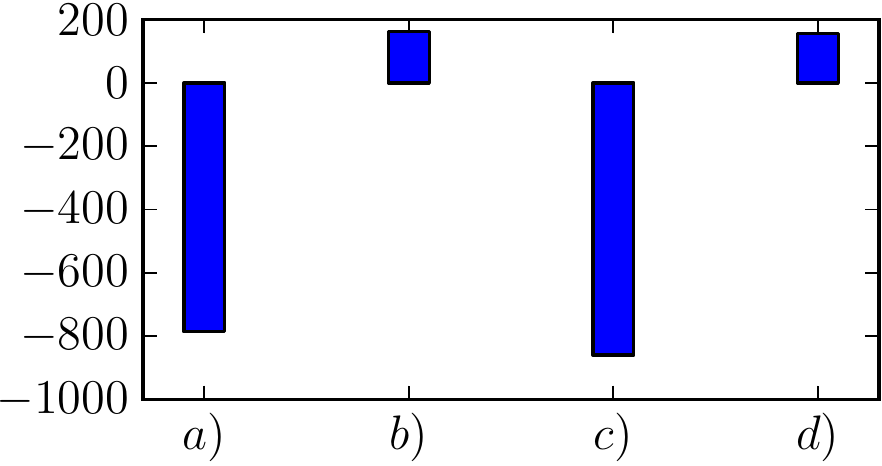}}
	\caption{The state of the network for the [2,2] game on non-periodic
	lattice (a)), A+B game on non-periodic lattice (b)), [2,2] game on
	periodic lattice (c)) and A+B game on periodic lattice (d)) after 1000
	iterations. The initial coin state is the W state. Figure e) shows
	the payoff for games A, B, [2,2] and A+B as a function of the number
	of players. Figure f) shows the average capital gain of an agent in
	the network after 1000 iterations.}\label{fig:W}
\end{figure}

Finally, we show results for the W state in Figure~\ref{fig:W}.
Figures~\ref{fig:W_0} and  \ref{fig:W_1} show the
results of the simulation for the non-periodic  network and
Figures~\ref{fig:W_2} and \ref{fig:W_3} show the results  for the periodic
network. The capital distribution in the case of the $A+B$ resembles the GHZ
case, whereas the distribution for the [2,2] is flat. This is
due to the fact that, the game does not show paradoxical behavior in the three
player case, as game B is a winning one. Nevertheless, the paradox is observed
for the four and five player A+B games as shown in Figure~\ref{fig:W_effect}. 
This leads to high average capitals of
the network in the case of the A+B game. As game [2,2] is a losing one for all
studied number of players, the average capitals of the networks are always
negative in this case as shown in Figure~\ref{fig:W_avg_cap}.

\section{Conclusions}\label{sec:conc}
We introduced a general scheme for executing quantum games on regular lattices,
which allows for the usage of entangled states and enables uniform analysis of
different scenarios.
To demonstrate the merits of the introduced scheme, we studied the
quantum Parrondo effect and the quantum prisoner's dilemma game in regular
lattices of agents.

The results for the prisoners dilemma game suggest that addition of the
quantum strategy $\Sigma$ to the classical strategies: cooperation ($C$) and
defection ($D$) gives a higher average capital gain of an agent than in the
classical case.

In the case of Parrondo paradox, the games showing paradoxical behavior were
modeled using quantum walks. We obtained results, showing that the
average capital of an agent in the lattice grows in the following setups of the
game:
\begin{itemize}
	\item $[2,2]$ game on a periodic lattice with the GHZ coin state,
	\item A+B game on a periodic lattice with the GHZ coin state,
	\item $[2,2]$ game on a non-periodic lattice with the GHZ coin state,
	\item A+B game on a non-periodic lattice with the GHZ coin state,
	\item A+B game on a periodic lattice with the W coin state,
	\item A+B game on a non-periodic lattice with the W coin state.
\end{itemize}
The above results show that entanglement is a necessary condition for the
lattice to gain capital as a whole.

\begin{acknowledgements}
Work by J.~S{\l}adkowski was supported by the Polish National Science Centre
under the project number DEC-2011/01/B/ST6/07197. Work by {\L}.~Pawela and J.
Miszczak was supported by the National Science Centre under the project number
DEC-2011/03/D/ST6/00413.
\end{acknowledgements}

\bibliography{parrondo_graph}
\bibliographystyle{apsrev}

\end{document}